\documentclass[%
reprint,
amsmath,amssymb,
aps,
]{revtex4-1}

\usepackage[caption=false]{subfig}
\usepackage{braket}
\usepackage[colorlinks=true,
linkcolor=blue]{hyperref}
\usepackage{graphicx}
\usepackage{dcolumn}
\usepackage{bm}

\newcommand{\degree}{$^{\circ}$}

\def\Vec#1{{\bf #1}}
\def\GVec#1{\mbox{\boldmath $#1$}}

\begin{document}
	
	\preprint{APS/123-QED}
	
	\title{Twist-angle dependence of the proximity spin-orbit coupling in graphene on transition-metal dichalcogenides}
	
	\author{Yang Li}
	\affiliation{Department of Physics, Tohoku University.}
	\author{Mikito Koshino}
	\affiliation{Department of Physics, Osaka University.}
	\date{\today}
	
\begin{abstract}
We theoretically study the proximity spin-orbit coupling in graphene on transition-metal dichalcogenides monolayer stacked with arbitrary twist angles. 
We find that the relative rotation greatly enhances the spin splitting of graphene, 
typically by a few to ten times compared to the non-rotated geometry,
and the maximum splitting is achieved around $20^\circ$.
The induced SOC can be changed from the Zeeman-type to the Rashba-type by rotation.
The spin-splitting is also quite sensitive to the gate-induced potential, and 
it sharply rises when the graphene's Dirac point is shifted toward the TMDC band.
The theoretical method does not need the exact lattice matching
and it is applicable to any incommensurate bilayer systems.
It is useful for the twist-angle engineering of a variety of van der Waals proximity effects.  
\end{abstract}
	
\maketitle
	
	
The physical properties of 2D material are generally sensitive to the interference with other materials placed in contact.
In recent years, a great deal of experimental and theoretical efforts have been made to explore the proximity-induced phenomena in van der Waals heterostructures consisting of different 2D crystals.\cite{geim2013van}
In particular, it was shown that the negligibly small spin-orbit coupling (SOC) of intrinsic graphene
can be significantly enhanced by superimposing on the surface of  transition-metal dichalcogenides (TMDC),
\cite{avsar2014spin, kaloni2014quantum, gmitra2015graphene, wang2015strong} 
and it is expected to be useful to realize spintronic manipulation in graphene.

In the studies on the proximity effect on 2D materials, however, the importance of  the relative lattice orientation 
has often been overlooked. The previous theoretical calculations of proximity spin-orbit effects of graphene/TMDC system
are limited to the non-rotated geometry. \cite{kaloni2014quantum, gmitra2015graphene, wang2015strong}
On the other hand, the sensitive dependence on the relative twist angle $\theta$ was noticed in various 2D hetrostructures,
and controlling $\theta$ is expected to be powerful means of manipulating their electronic properties. 
\cite{carr2017twistronics,palau2018twistable}
In graphene on hexagonal BN system, for instance, the moir\'{e} interference pattern gives rise to the formation of the secondary Dirac points 
and the miniband structure. \cite{kindermann2012zero, wallbank2013generic, mucha2013heterostructures, jung2014ab, moon2014electronic,dean2013hofstadter,ponomarenko2013cloning,hunt2013massive,yu2014hierarchy}  
The twisted bilayer graphene also exhibits the dramatic angle-dependent phenomena, such as the flat band formation
\cite{lopes2007graphene,mele2010commensuration,trambly2010localization,shallcross2010electronic,morell2010flat,bistritzer2011moire,moon2012energy,de2012numerical} and the emergent superconductivity. \cite{cao2018unconventional,cao2018mott}
For graphene/TMDC hetrostructure,
the twist-angle dependent band structure was theoretically simulated for several commensurate angles
by the density functional theory (DFT) \cite{wang2015electronic,di2017angle, hou2017robust},
and it is also experimentally probed. \cite{jin2015tuning, pierucci2016band,du2017h,lu2017moire}
However, the $\theta$-dependence of spin-orbit coupling induced on graphene remains still unclear.
It is generally hard to consider arbitrary twist angles in the DFT calculation, because it requires
exact lattice matching to have a finite unit cell.

\begin{figure}
	\includegraphics[width=0.9 \hsize]{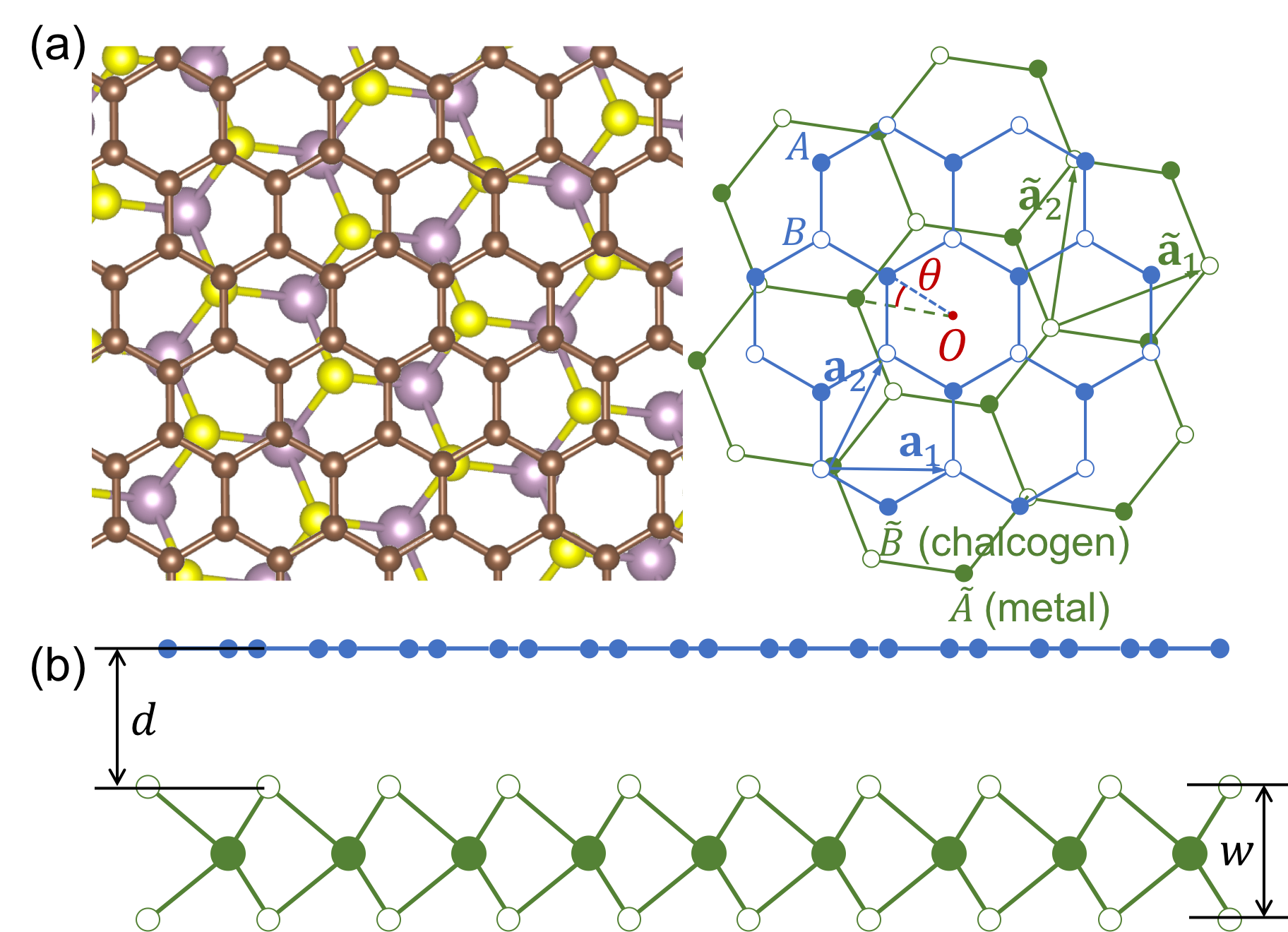}
	\caption{
(a)Top view and (b) the side view of graphene on TMDC monolayer with twist angle $\theta$.
}
\label{fig:atomic structure}
\end{figure}

In this letter, we theoretically study the proximity SOC effect in graphene-TMDC heterostructures 
with arbitrary twist angles $\theta$, and reveal the angle dependence of SOC for various different TMDCs.
Using the tight-binding model and the perturbational approach, which do not need the commensurate lattice matching,
we obtain the effective Hamiltonian of graphene as a continous function of $\theta$.
We find that the relative rotation greatly enhances the spin splitting, 
typically by a few to ten times compared to the non-rotated geometry $(\theta=0)$,
and the maximum splitting is achieved around $\theta \sim 20^\circ$.
We also show that the induced SOC is composed of the Zeeman-like term and the rotated Rashba-like term,
and the relative magnitude can be controlled by rotation.
Finally, we demonstrate that the spin-splitting is quite sensitive to the relative band energy between graphene and TMDC,
and it sharply rises when the graphene's Dirac point is shifted toward TMDC band by applying the gate voltage. 
The theoretical method proposed here is applicable to any incommensurate bilayer systems where the DFT calculation cannot be used, and therefore it considerably extends the applicability of the theoretical framework to a wide variety of
van der Waals heterostructures. 



We consider monolayer graphene placed on the top of a TMDC monolayer. 
Graphene and TMDC are two dimensional honeycomb lattices with different lattice periods,
$a_G = 2.46$\AA \, for graphene and $a_T$ for TMDC given in table \ref{tab:table1}.
We define the stacking geometry starting from non-rotated arrangement with parallel bond directions,
and then rotating TMDC by the twist angle $\theta$ around the common center of hexagon as in Fig.\ \ref{fig:atomic structure}(a).
The lattice structure has the $C_3$ (120$^\circ$) rotational symmetry with respect to the rotation center.
We neglect the degree of freedom of the in-plane parallel translation between TMDC and graphene,
since in an incommensurate system it can always be incorporated with the shift of the origin. \cite{koshino2015interlayer}

The lattice vectors of graphene are then given by $\mathbf{a}_1 = a_G(1, 0)$ and 
$\mathbf{a}_2 = a_G(1/2, \sqrt{3}/2)$, and 
those of TMDC are by $\tilde{\mathbf{a}}_1 = R a_T(1, 0)$ and $\tilde{\mathbf{a}}_2 = R a_T(1/2, \sqrt{3}/2)$,
 where $R= R(\theta)$ is the rotation matrix. 
 The unit cell area is  $S = |\mathbf{a}_1 \times \mathbf{a}_2|$ and $\tilde{S} = |\tilde{\mathbf{a}}_1 \times \tilde{\mathbf{a}}_2|$  for graphene and TMDC, respectively. 
The reciprocal lattice vectors, $\mathbf{a}^*_1, \mathbf{a}^*_2, \tilde{\mathbf{a}}^*_1, \tilde{\mathbf{a}}^*_2$, are defined by $\mathbf{a}_i \cdot \mathbf{a}^*_j = \tilde{\mathbf{a}}_i \cdot \tilde{\mathbf{a}}^*_j = 2 \pi \delta_{ij}$. 
We define $d$ as the distance between the graphene layer and the top chalcogen layer,
and $w$ as the distance between top and bottom chalcogen layers. The values of $d$ and $w$
depend on TMDCs as shown in table \ref{tab:table1}.

We model graphene by the tight-binding model of carbon $p_z$ orbitals, where the sublattice is labeled as
$X= p_z^{A}, p_z^{B}$ for A and B sites, respectively.
For TMDC, we adopt the tight-binding model including three $p$ orbitals for a chalcogen atom and five $d$ orbitals for a transition metal atom \cite{fang2015ab}.
The orbitals in a TMDC unit cell is labeled by $\tilde{X}= d_{z^2},d_{xy},d_{x^2-y^2},d_{xz},d_{yz}, p_x^{t},  p_y^{t},  p_z^{t}, p_x^{b}, p_y^{b}, p_z^{b}$,
where $t$ and $b$ represent top and bottom chalcogen layers.
The positions of the orbitals are given by
\begin{align}
&\Vec{R}_{X}=n_1\Vec{a}_{1}+n_2\Vec{a}_{2}+\GVec{\tau}_X
\quad (\mbox{graphene}), \nonumber\\
&\Vec{R}_{\tilde{X}}=\tilde{n}_1\tilde{\Vec{a}}_{1}+\tilde{n}_2\tilde{\Vec{a}}_{2}+\GVec{\tau}_{\tilde{X}}
\quad (\mbox{TMDC}),
\end{align}
where $n_i$ and $\tilde{n}_i$ are integers,
and $\GVec{\tau}_X$ and $\GVec{\tau}_{\tilde{X}}$ are the sublattice position
inside the unit cell. Specifically, they are expressed as
$\boldsymbol{\tau}_{p_z^{A}} =  - \boldsymbol{\tau}_1$, 
$\boldsymbol{\tau}_{p_z^{B}} = \boldsymbol{\tau}_1 $ for graphene,
and 
$\boldsymbol{\tau}_{\tilde{X}}  =  -\tilde{\boldsymbol{\tau}}_1 - (d+w/2)  \Vec{e}_z$ for the transition metal $d$-orbitals and
$\boldsymbol{\tau}_{\tilde{X}}  = \tilde{\boldsymbol{\tau}}_1 -d \Vec{e}_z, \tilde{\boldsymbol{\tau}}_1 -(d+w) \Vec{e}_z$ 
for the top and bottom charcogen $p$-orbitals, respectively,
where $\boldsymbol{\tau}_1=(-\mathbf{a}_1 + 2 \mathbf{a}_2)/ 3$
and $\tilde{\boldsymbol{\tau}}_1=(-\tilde{\mathbf{a}}_1 + 2 \tilde{\mathbf{a}}_2)/ 3$.

The Hamiltonian is spanned by the Bloch bases,
\begin{align}
	& |\Vec{k},X,s\rangle = 
	\frac{1}{\sqrt{N}}\sum_{\Vec{R}_{X}} e^{i\Vec{k}\cdot\Vec{R}_{X}}
	|\Vec{R}_{X}, s \rangle\quad (\mbox{graphene}), \nonumber\\
	& |\tilde{\Vec{k}},\tilde{X},\tilde{s}\rangle = 
	\frac{1}{\sqrt{\tilde{N}}}\sum_{\Vec{R}_{\tilde{X}}} e^{i\tilde{\Vec{k}}\cdot\Vec{R}_{\tilde{X}}}
	|\Vec{R}_{\tilde{X}},\tilde{s}\rangle \quad (\mbox{TMDC}),
	\label{eq_bloch_base}
\end{align}
where $s, \tilde{s}$ are the spin indexes, $\Vec{k}$ and $\tilde{\Vec{k}}$ are the two-dimensional
Bloch wave vectors parallel to the layer,
and $N = S_{\rm tot}/S$ and $\tilde{N}=S_{\rm tot}/\tilde{S}$ are the number of unit cells of TMDC and graphene, respectively, 
in the total system area $S_{\rm tot}$. 

\begin{table}[b]
		\caption{\label{tab:table1}%
		        List of parameters for TMDCs and graphene-TMDC bilayers used in this work (see the text).
			}
		\begin{ruledtabular}
			\begin{tabular}{ccccc}
				&
				\textrm{MoS$_2$}&
				\textrm{MoSe$_2$}&
				\textrm{WS$_2$}&
				\textrm{WSe$_2$}\\
				\colrule
				$a_T$(\AA) & 3.18\cite{fang2015ab} & 3.32\cite{fang2015ab} & 3.18\cite{fang2015ab} & 3.32\cite{fang2015ab}\\
				$w$(\AA)  & 3.13\cite{fang2015ab} & 3.34\cite{fang2015ab} & 3.14\cite{fang2015ab} & 3.35\cite{fang2015ab}\\
				$d$(\AA)  & 3.37\cite{fang2015ab} & 3.41\cite{ma2011first} & 3.41\cite{kaloni2014quantum} & 3.42\cite{kaloni2014quantum} \\
				$E_T-E_G$(eV) &  0.02\cite{gmitra2015graphene} & 0.6\cite{ma2011first} &  0.12\cite{kaloni2014quantum} & 1.06\cite{agnoli2018unraveling, kaloni2014quantum}
			\end{tabular}
		\end{ruledtabular}
\end{table}

\begin{figure}
	\includegraphics[width=1.\hsize]{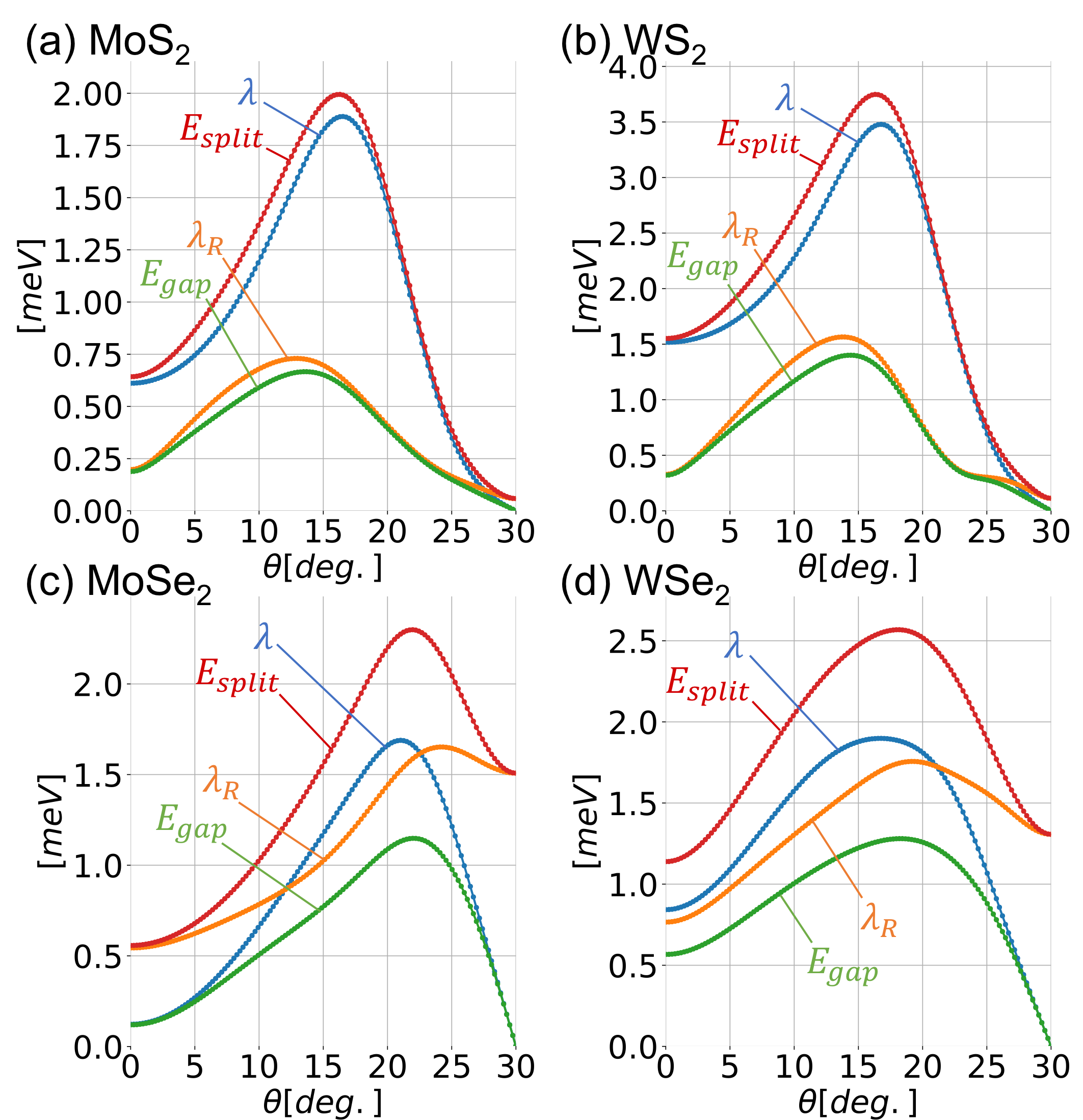}
		\caption{Spin-orbit parameters $\lambda$, $\lambda_R$, 
the central energy gap $E_{\rm gap}$ and the spin splitting $E_{\rm split}$, as a function of the twist angle $\theta$
in graphene-TMDC bilayers.}
		\label{fig_lambda}
	\end{figure}


The total tight-binding Hamiltonian is expressed as
$H = H_{\rm G} + H_{\rm T} + H_{\rm int},$
where $H_{\rm G}$ and $H_{\rm T}$ are the Hamiltonian for the intrinsic graphene monolayer and TMDC monolayer, respectively,
and $H_{\rm int}$ is for the coupling between graphene and TMDC.
For $H_{\rm T}$, we adopt the hopping parameters based on the first principles calculation \cite{fang2015ab}
where the spin-orbit coupling is included by on-site $\Vec{L}\cdot\Vec{S}$ term for each atom.
The on-site energy of the TMDC atoms relative to the carbon atoms
is extracted from the relative energy $E_T-E_G$ from the graphene Dirac point to TMDC conduction band edges in 
the first principles calculations\cite{gmitra2015graphene, agnoli2018unraveling, ma2011first, ma2011first, kaloni2014quantum},
which are listed in Table \ref{tab:table1}. 


For the interlayer interaction, we assume that the transfer integral from $\Vec{R}_{X}$ to $\Vec{R}_{\tilde{X}}$
is expressed as $-T_{\tilde{X}X}(\Vec{R}_{\tilde{X}} - \Vec{R}_{X})$,
with the Slater-Koster parameterization \cite{slater1954simplified} and the exponential decay in the distance.
Here the hopping amplitude and the decay length are determined to fit the first principles calculations.
The detailed method is described in the supplementary materials.
The coupling between the Bloch state of graphene and that of TMDC is then given by  \cite{bistritzer2011moire,koshino2015interlayer}
	\begin{eqnarray}
		&& \braket{\tilde{\mathbf{k}}, \tilde{X}, \tilde{s} | H_{\rm int} | \mathbf{k}, X, s} = \nonumber\\ 
		&& -\sum_{\mathbf{G}, \tilde{\mathbf{G}}}  t_{\tilde{X}X}(\mathbf{k} + \mathbf{G}) e^{-i \mathbf{G} \cdot \boldsymbol{\tau}_X + i \tilde{\mathbf{G}} \cdot \boldsymbol{\tau}_{\tilde{X}}} \delta_{\mathbf{k} + \mathbf{G}, \tilde{\mathbf{k}} + \tilde{\mathbf{G}}} \delta_{\tilde{s}s} .
		\label{eq:interlayer coupling hamiltonian}
	\end{eqnarray}
Here $\Vec{G}=m_1\Vec{a}^*_1+m_2\Vec{a}^*_2$ and
$\tilde{\Vec{G}}=\tilde{m}_1\tilde{\Vec{a}}^*_1+\tilde{m}_2\tilde{\Vec{a}}^*_2$
are reciprocal lattice vectors of graphene and TMDC, respectively,
${t}_{\tilde{X}X}(\Vec{q})$ is the in-plane Fourier transform
of the transfer integral defined by
\begin{eqnarray}
{t}_{\tilde{X}X}(\Vec{q}) = 
\frac{1}{\sqrt{S\tilde{S}}} \int
T_{\tilde{X}X}(\Vec{r}+ z_{\tilde{X}X}\Vec{e}_z) 
e^{-i \Vec{q}\cdot \Vec{r}} d^2r,
\label{eq_ft}
\end{eqnarray}
where $z_{\tilde{X}X} = (\GVec{\tau}_{\tilde{X}}-\GVec{\tau}_{X})\cdot\Vec{e}_z$.

\begin{figure}
	\includegraphics[width=1.\hsize]{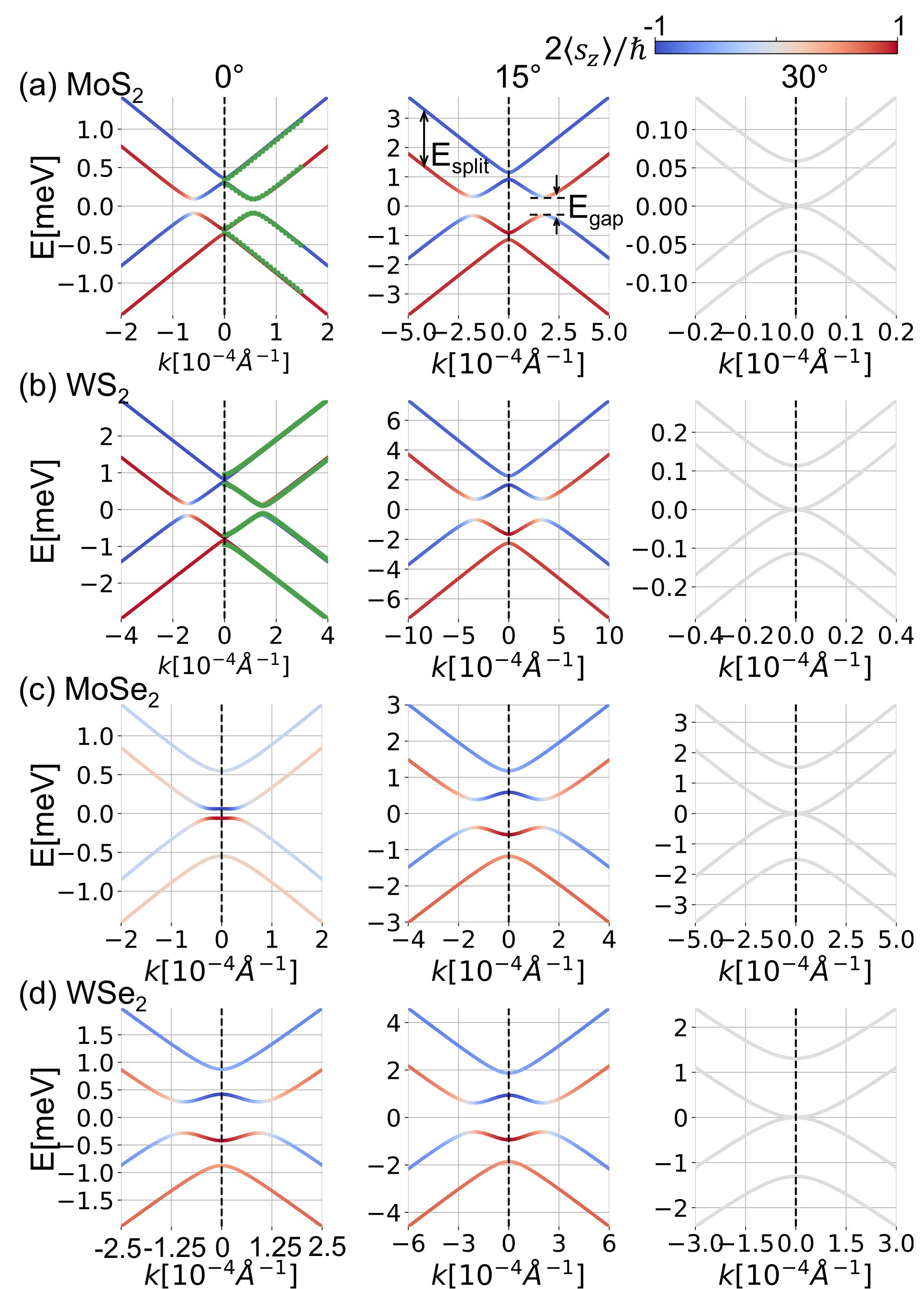}
\caption{Band structures for graphenes on (a)MoS$_2$, (b)WS$_2$, (c)MoSe$_2$ and (d)WSe$_2$ 
at the twist angles $\theta =0^\circ, 15^\circ$, and $30^\circ$, where color indicates the expectation value
of $s_z$. In the band plots of MoS$_2$ and WS$_2$ at $\theta =0^\circ$, 
the dotted green line indicates the DFT calculations.}
\label{fig_band_all}
\end{figure}

	
The Hamiltonian of graphene including the TMDC proximity effect 
can be obtained by the second order perturbation as $H_{\rm eff} (\Vec{k})= H_G (\Vec{k})  + V_{\rm eff} (\Vec{k})$, where
\begin{align}
& [V_{\rm eff} (\Vec{k})]_{X's',Xs}   =
\nonumber\\
&\quad  
\sum_{\tilde{n},\tilde{\Vec{k}}} 
 \frac{
\braket{\mathbf{k}, X', s' | H_{\rm int} | \tilde{n},\tilde{\Vec{k}}}
\braket{\tilde{n},\tilde{\Vec{k}} | H_{\rm int} | \mathbf{k}, X, s}
 }
 {E_G - E_{\tilde{n},\tilde{\Vec{k}}}}.
\label{eq_V_eff}
\end{align}
Here $E_G$ is the energy of the graphene's Dirac point,
and $E_{\tilde{n},\tilde{\Vec{k}}}$ and $|\tilde{n},\tilde{\Vec{k}}\rangle$ are the eigen energy and eigen state of $H_{\rm T}$, respectively,
with the band index $\tilde{n}$ (including the spin degree of freedom) and the Bloch vector $\tilde{\Vec{k}}$.
Note that $|\tilde{n},\tilde{\Vec{k}}\rangle$ is written as a linear combination of $|\tilde{\mathbf{k}}, \tilde{X}, \tilde{s} \rangle$ of the same $\tilde{\Vec{k}}$.
The summation over $\tilde{\Vec{k}}$ in Eq.\ (\ref{eq_V_eff})
is taken according to the condition Eq.\ (\ref{eq:interlayer coupling hamiltonian}).

The low-energy Hamiltonian is obtained by expanding  $H_{\rm eff} (\Vec{k})$ 
around the valley center $\mathbf{K}_\xi \equiv -\xi(2\mathbf{a}^*_1 + \mathbf{a}^*_2) / 3$,
where $\xi = \pm 1$ is the valley index.
Within the linear term, $H_{\rm G}$ is approximated by $H^{(\xi)}_G(\Vec{k}) = -\hbar v (\mathbf{k} - \mathbf{K}_\xi) \cdot (\xi \sigma_x, \sigma_y)$,
where $v$ is the band velocity of graphene, and $\sigma_x$ and $\sigma_y$ are Pauli matrices for the sublattice space
$X= p_z^{A}, p_z^{B}$. \cite{ando2009electronic}.
For the proximity SOC term, we only take the zero-th order $V_{\rm eff}(\Vec{K}_\xi) \equiv V^{(\xi)}_{\rm eff}$.
Now that the transfer integral $T_{\tilde{X}X}(\Vec{R})$ attenuates exponentially and so does its Fourier transform $t_{\tilde{X}X}(\mathbf{q})$, 
it suffices to keep only a few $\tilde{\Vec{k}}$'s in the summation of Eq.\ (\ref{eq_V_eff}).
For  $\Vec{k}=\Vec{K}_\xi$, the dominant contribution comes from three points,
$\tilde{\mathbf{k}} = \mathbf{K}_\xi  + \xi \tilde{\Vec{a}}^*_1, \mathbf{K}_\xi  + \xi (\Vec{a}^*_1 + \tilde{\Vec{a}}^*_2), 
\mathbf{K}_\xi  + \xi (\Vec{a}^*_1 + \Vec{a}^*_2 -  \tilde{\Vec{a}}^*_1 - \tilde{\Vec{a}}^*_2)$,
while the effect of other $\tilde{\Vec{k}}$'s are negligibly small.
In this way, the effective proximity potential can be obtained by considering TMDC Bloch states at only three wave points,
and the corresponding computing cost is considerably low.

We can show that $V^{(\xi)}_{\rm eff}$ can be written as,
\begin{equation}
V^{(\xi)}_{\rm eff}  = \frac{\lambda}{2} \xi s_z +  \frac{\lambda_R}{2} e^{-i \phi s_z/2}(\xi \sigma_x s_y - \sigma_y s_x)e^{i \phi s_z/2},
\label{eq:V_eff_short}
\end{equation}
where $s_i \, (i=x,y,z)$ is the Pauli matrix for spin. 
It is explicitly written in a matrix form,
\begin{eqnarray}
&V^{(+)}_{\rm eff}
          = \left( \begin{array}{cccc}
           \lambda/2 & & & \\
            &  \lambda/2 & -i \lambda_R e^{-i\phi} & \\
            & i \lambda_R e^{i\phi}  & - \lambda/2 & \\
            & & & - \lambda/2
\end{array} \right),
\nonumber\\
&V^{(-)}_{\rm eff}          = \left( \begin{array}{cccc}
           - \lambda/2 & & & i \lambda_R e^{-i\phi}  \\
            & - \lambda/2 &  & \\
            && \lambda/2 & \\
            -i \lambda_R e^{i\phi} & & &  \lambda/2
\end{array} \right),
\label{eq: effective hamiltonian-2}
	\end{eqnarray}
where the bases are arranged by order of $(X,s)=(A, \uparrow)$, $(B, \uparrow)$, $(A, \downarrow)$ and $(B, \downarrow)$. 
The difference in the diagonal elements $\lambda$ leads to the spin splitting between spin up and spin down, 
and the off-diagonal term $\lambda_R$ mixes the different spins.
The term with $\lambda_R$ is similar to the Rashba spin-orbit coupling \cite{kane2005quantum, wang2015strong}
but here the spin axis can be rotated by an angle $\phi$ on $xy$-plane.
The energy gap at the charge neutral point is given by $E_{\rm gap} = |\lambda\lambda_R|/(\lambda^2+\lambda_R^2)^{1/2}$.
The spin splitting width in large $k$ is given by $E_{\rm split} = (\lambda^2+\lambda_R^2)^{1/2}$.
The effective Hamiltonian $H^{(\xi)}_{\rm eff} = H^{(\xi)}_G(\Vec{k}) + V^{(\xi)}_{\rm eff}$ 
is formally equivalent with that of the asymmetric bilayer graphene, \cite{mccann2006asymmetry, mccann2013electronic}
where the spin up and down correspond to layer 1 and 2, respectively,
and $\lambda_R$ and $\lambda$ to the interlayer coupling and the interlayer asymmetric potential, respectively.

The form of Eq.\ (\ref{eq:V_eff_short}) is forced by the symmetry of the system.
The terms in Eq.\ (\ref{eq:V_eff_short}) are generally allowed in the time reversal symmetry $\cal{T}$ and  the $C_3$ symmetry. 
Actually, the term proprotional to $\sigma_z$ (different on-site energies at $A$ and $B$ sites)
is also possible under $\cal{T}$ and $C_3$ \cite{wang2015strong}, 
while it is prohibited by the incommesurability between graphene and TMDC, as explained in Supplementary Material.
An additional space symmetry imposes a constraint on $V^{(\xi)}_{\rm eff}$.
At $\theta =0$, the reflection symmetry $R_x:(x,y,z)\to (-x,y,z)$ requires $e^{i\phi}$ is real.
At $\theta =30^\circ$, the reflection symmetry $R_y:(x,y,z)\to (x,-y,z)$ requires real $e^{i\phi}$
and also $\lambda = 0$, i.e., the SOC is dominated by the Rashba term.
The detailed argument of the symmetry consideration is presented in Supplementary Material.



\begin{figure}
\includegraphics[width=\hsize]{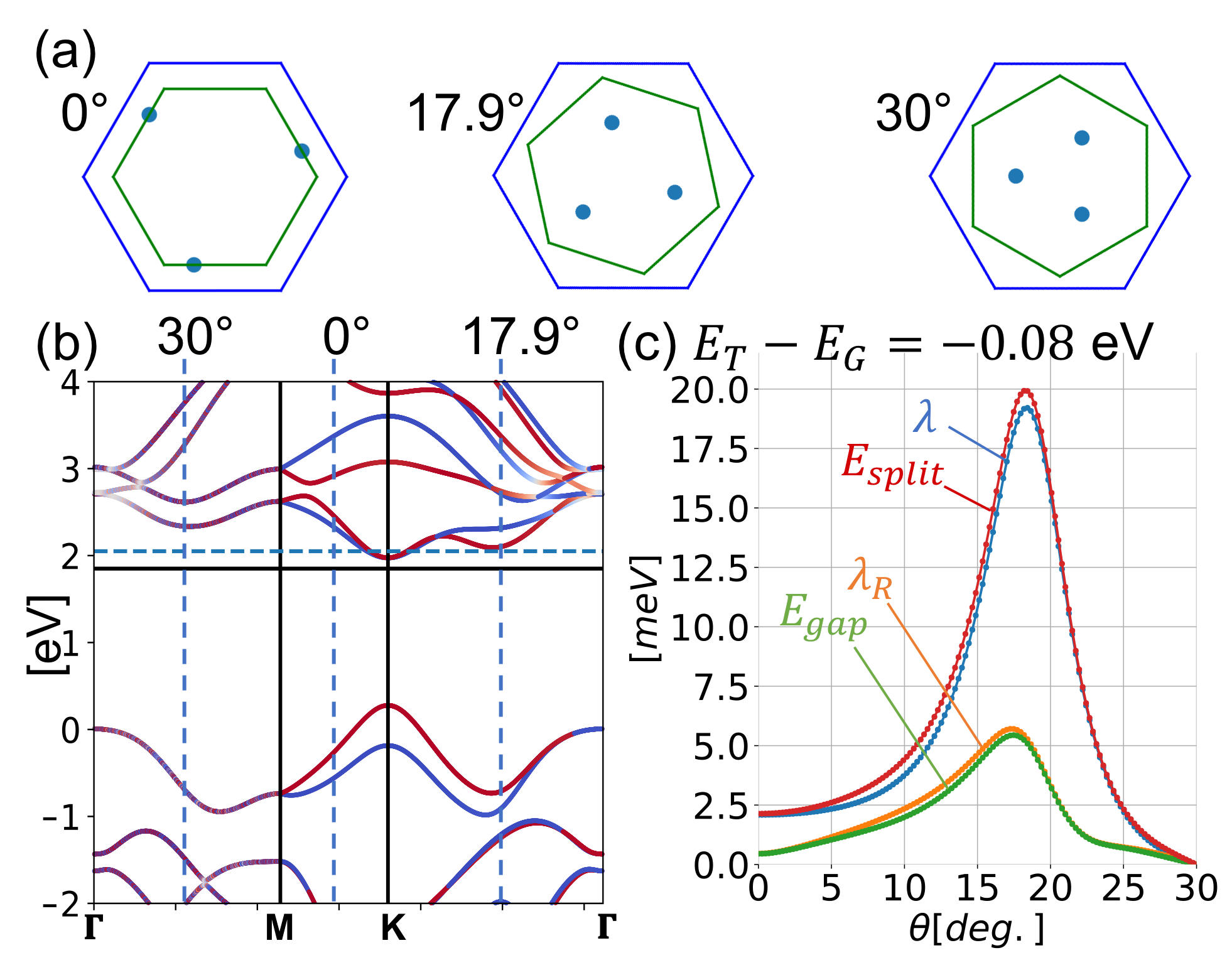}
\caption{(Top) Position of three dominant $\tilde{\mathbf{k}}$ points 
for $\xi = +$,  in WS$_2$ with $\theta = 0^\circ, 17.9^\circ$, and $30^\circ$.
Blue (green) hexagon represents the first Brillouin zone of graphene (WS$_2$).
(Bottom left) Band structure of WS$_2$, where the vertical dashed lines indicate the $\tilde{\mathbf{k}}$ for the three rotation angles, and the black horizontal line is the energy of graphene's Dirac cone without gate voltage. 
(Bottom Right) Plot similar to Fig.\ \ref{fig_lambda},
calculated for WS$_2$ with $E_T-E_G = -0.08$ eV, indicated by the blue dashed horizontal line in (b)}
\label{fig_WS2_band}
\end{figure}


We numerically calculate $V^{(\xi)}_{\rm eff}$ for MoS$_2$,  WS$_2$, MoSe$_2$ and WSe$_2$.
Figure \ref{fig_lambda} summarizes the results, where $\lambda$, $\lambda_R$, 
the central energy gap $E_{\rm gap}$ and the spin splitting $E_{\rm split}$ are plotted against the twist angle $\theta$.
In Fig.\ \ref{fig_band_all}, we present the band structures for each system
at the rotation angles $\theta =0^\circ, 15^\circ$, and $30^\circ$.
In the band plots of MoS$_2$ and WS$_2$ at $\theta =0^\circ$, 
the dotted green line indicates the first DFT calculations, from which we extract the interlayer hopping parameters.
For the DFT calculation, we assume the approximate commensurate lattice structure of which unit cell is
comprised of 3$\times$3 supercell of MoS$_2$ and 4$\times$4 of graphene,
and use Quantum Espresso\cite{giannozzi2009quantum, giannozzi2017advanced} 
with the generalized gradient approximation\cite{perdew1996generalized}. 
We can see that the effective model well reproduces the DFT band structure.

For the angle dependence, we find that  $\lambda$ and $\lambda_R$ are greatly enhanced by rotation,
and they take the maximum around $\theta \sim 20^\circ$.
For WS$_2$, in particular, the maximum splitting is about 5 times as large as that of 0$^\circ$.
At 30\degree, the parameter $\lambda$ vanishes and the $V_{\rm eff}$ is dominated by
$\lambda_R$ as expected. There the band structure is formally equivalent to the symmetric AB-stacked bilayer graphene, 
and the expectation value of spin lies on the $xy$-plane.

The enhancement of the spin-splitting near $20^\circ$ can be understood by considering 
the second-order process,  Eq.\ (\ref{eq_V_eff}).
The amplitude of $V_{\rm eff}$ is related to the spin splitting of the TMDC
bands at $\tilde{\mathbf{k}}$ points which are hybridized with the graphene's Dirac point.
Figure \ref{fig_WS2_band}(a) illustrates the positions of the three dominant $\tilde{\mathbf{k}}$'s for $\xi = +$,  
in WS$_2$ with $\theta = 0^\circ, 17.9^\circ$, and $30^\circ$.
Figure \ref{fig_WS2_band}(b) presents the band structure of WS$_2$
with the vertical dashed lines indicating the $\tilde{\mathbf{k}}$'s for the three rotation angles.
Now the lowest valence band of TMDC makes the greatest contribution to $V_{\rm eff}$,
as it is the closest to the graphene's Dirac point energy (black horizontal line), leading to a small denominator in Eq.\ (\ref{eq_V_eff}).
We can see that the $\tilde{\mathbf{k}}$ point for $17.9^\circ$ happens to be very close to 
the $Q$-valley, where the magnitude of the spin splitting much greater than in other angles.
This qualitatively explains the sharp rise of $\lambda$ and $\lambda_R$ around $20^\circ$.
Actually, the spin splitting can be even enhanced by shifting the relative energy between graphene and TMDC.
Figure \ref{fig_WS2_band}(c) plots the angle dependence of the spin-splitting of WS$_2$
with $E_T-E_G =  -0.08$ eV, where the graphene's Fermi energy (blue horizontal line) hits the bottom of the Q-valley.
Although the Fermi energy is just a little higher than in Fig.\ \ref{fig_lambda}(b),
the maximum spin splitting sharply increases to 20 meV, about 10 times as big as in $\theta =0$,
because the denominator in Eq.\ (\ref{eq_V_eff}) becomes very small.
This suggests that tuning of the spin-orbit coupling would be possible using the external gate voltage.

	
Finally, the graphene under the proximity potential has the non-zero valley Hall conductivity when the Fermi energy lies in the central gap.
The Hall conductivity of each valley sector can be calculated using the Berry curvature as
\begin{align}
\sigma^{(\xi)}_{xy} = \frac{e^2}{h} \sum_{n \in {\rm occ.}} \int \frac{d^2\Vec{k}}{2\pi} \nabla_\Vec{k}\times \Vec{a}_n(\Vec{k})
= -\frac{e^2}{h} \, \xi \, {\rm sgn}(\lambda),
\end{align}
where $\Vec{a}_n(\Vec{k}) = -i \langle u_{n\Vec{k}}| \nabla_\Vec{k} | u_{n\Vec{k}}\rangle$
is the Berry connection, $u_{n\Vec{k}}$ is the Bloch function (eigenvector of $H^{(\xi)}_{\rm eff}$) of the band $n$,
and occ. stands for the occupied valence bands ($n=1,2$). 
As a result, the valley Hall conductivity becomes $\sigma^{(+)}_{xy} - \sigma^{(-)}_{xy} = -(2e^2/h){\rm sgn}(\lambda)$.

	
To conclude,  we have studied the proximity spin-orbit interaction in graphene-TMDC bilayers stacked with arbitrary twist angles.
By using the perturbational approach based on the tight-binding model,
we derived the effective Hamiltonian of graphene as a continuous function of the twist angle $\theta$,
and found that the magnitude of SOC is greatly enhanced by the rotation.
We also show that the SOC sharply rises when the graphene's Dirac point is shifted toward TMDC band, by applying the gate voltage. 
The theoretical method proposed here does not need the exact lattice matching,
so that it is applicable to any incommensurate bilayer systems which cannot be treated by the DFT calculation.
It would be useful for the twist-angle engineering of a wide variety of van der Waals proximity effects, 
including ferromagnetism and superconductivty.
M. K. acknowledges the financial support of JSPS KAKENHI Grant No. JP17K05496.

	\bibliography{reference}
	
\end{document}